\documentclass[twocolumn,10pt]{article} 

\usepackage[square,numbers,sort&compress,comma]{natbib}

\usepackage{amsmath}
\usepackage{amssymb}
\usepackage{caption}
\usepackage{graphicx}
\usepackage{latexsym}
\usepackage{times}
\usepackage[pagewise]{lineno}
\usepackage{subeqnarray}
\usepackage{caption}
\usepackage{subcaption}
\usepackage{overpic}
\usepackage{floatrow}
\usepackage{subfig}
\usepackage[export]{adjustbox}
\usepackage{float}
\usepackage{stfloats}
\usepackage{arydshln}
\usepackage{xcolor}
\usepackage{booktabs}

\makeatletter
\newcommand\rlarrows{\mathop{\operator@font \rightleftarrows}\nolimits}
\makeatother
 \def\0vec{{\mbox{\boldmath$0$}}}

\topmargin - 12pt 
\renewenvironment{abstract}%
              {
               \small
               {\bfseries \abstractname}
               \par
               \vspace{10pt}
              }

\renewcommand\abstractname{Abstract}

\newcommand{\nomenclature}
              [1]
              {
               \bgroup
               \flushleft
               \small\bf
               #1
               \par
               \egroup
              }

\renewcommand{\section}
              [1]
              {
               \bgroup
               \flushleft
               \small\bf
               \refstepcounter{section}
               \arabic{section}. #1
               \par
               \egroup
              }

\renewcommand{\subsection}
              [1]
              {
               \bgroup
               \flushleft
               \small\em
               \refstepcounter{subsection}
               \arabic{section}.
               \arabic{subsection}. #1
               \par
               \egroup
              }

\renewcommand{\subsubsection}
              [1]
              {
               \bgroup
               \flushleft
               \small\em
               \refstepcounter{subsubsection}
               \arabic{section}.
               \arabic{subsection}.
               \arabic{subsubsection}. #1
               \par
               \egroup
              }

  \newcommand{\acknowledgement}
              [1]
              {
               \bgroup
               \flushleft
               \small\bf
               #1
               \par
               \egroup
              }

  \newcommand{\sectionbib}
              [1]
              {
               \bgroup
               \flushleft
               \small\bf
               #1
               \par
               \egroup
              }

\begin{document}

\title{\LARGE Linear stability and resolvent analyses 
 of a bluff-body stabilized flame with conjugate heat transfer}

\author{{\large Lu Chen$^{a,b}$,  Wai Lee Chan$^c$, Yu Lv$^{a,d,*}$}\\[10pt]
        {\footnotesize \em $^a$State Key Laboratory of Nonlinear Mechanics, Institute of Mechanics, Chinese Academy of Sciences, Beijing 100080,
China}\\[-5pt]
        {\footnotesize \em $^b$Department of Mechanical Engineering, National University of Singapore,
117575, Singapore} 
        \\[-5pt]
        {\footnotesize \em $^c$ School Mechanical and Aerospace Engineering, Nanyang Technological University, Singapore, 639798, Singapore
 Singapore}
        \\[-5pt] 
        {\footnotesize \em $^d$School of Engineering Sciences, University of Chinese Academy of Sciences, Beijing 100049, China}\\[-5pt]
        }

\date{}


\small
\baselineskip 10pt


\twocolumn[\begin{@twocolumnfalse}
\vspace{50pt}
\maketitle
\vspace{40pt}
\rule{\textwidth}{0.5pt}
\begin{abstract} 
Conjugate heat transfer is a challenging fluid-structure coupling problem that can significantly 
influence flame stabilization and thermoacoustic instabilities.  
To properly capture combustion phenomena that involve conjugate heat transfer, careful modeling 
of chemical reactions in the fluid domain and heat transfer in the solid body is necessary and 
remains an active research topic. To this end, 
we derived a strongly-coupled method with a monolithic weak formulation to investigate the conjugate heat transfer between an anchored flame and a thermal conductive cylinder by means of linear stability analysis and resolvent analysis. We conduct parameter continuation with the Damk\"{o}hler number to construct a bifurcation diagram and identify multiple baseflow states, including blow-off, anchored flame, and flashback. Linear stability analysis reveals the presence of a single unstable, non-oscillatory eigenmode for the base states on the anchored flame branch. This eigenmode plays a pivotal role in driving the bifurcation. Subsequently, resolvent analysis is performed to examine the amplification behavior of the fluid-solid coupled system under external forcing, showing that heat fluctuations are maximized when heat transfer between the fluid and solid is minimized.
\end{abstract}
\vspace{10pt}
\parbox{1.0\textwidth}{\footnotesize {\em Keywords:} Conjugate heat transfer; Flame dynamics; Bifurcation; Linear analysis; Resolvent analysis}
\rule{\textwidth}{0.5pt}
\vspace{180pt}

*Corresponding author\\
Email: lvyu@imech.ac.cn (Y. Lv)

\end{@twocolumnfalse}] 

\clearpage

\twocolumn[\begin{@twocolumnfalse}

\centerline{\bf Information for Colloquium Chairs and Cochairs, Editors, and Reviewers}

\vspace{20pt}

\vspace{20pt}

{\bf 1) Novelty and Significance Statement}
\vspace{10pt}

\vspace{10pt}

This study presents a novel, strongly-coupled monolithic numerical framework for performing linear stability and resolvent analyses of bluff-body flames with conjugate heat transfer effects, enabling a unified approach to investigate both intrinsic and forced combustion dynamics. By systematically varying an effective Damk\"{o}hler number, the solver captures multiple flame base states, including flashback, anchored, and blow-off conditions, offering a comprehensive framework for exploring flame behavior under varying operational regimes. Stability analysis reveals the existence of a low-frequency unstable mode for flames anchored on the cylinder surface, highlighting a critical instability mechanism. Resolvent analysis further demonstrates that the flame’s response to external forcing is highly sensitive to its base state. This work advances the understanding of flame dynamics and provides a robust tool for analyzing thermoacoustic instabilities in combustion systems.

\vspace{20pt} 

{\bf 2) Author Contributions}
\vspace{10pt}

\begin{itemize}

  \item{LC: Data curation; Formal analysis; Investigation; Methodology; Software; Validation; Visualization; 
   Writing -- original draft; Writing -- review \& editing }
  \item{WLC: Funding acquisition; Supervision; Writing -- review \& editing }
  \item{YL: Conceptualization; Formal analysis; Funding acquisition; Investigation; Methodology; Project administration; Resources; 
   Supervision; Validation; Visualization; Writing -- original draft; Writing -- review \& editing}
  
\end{itemize}

\vspace{10pt}

{\bf 3) Manuscript Length}
\vspace{10pt} 

The authors prefer eight-page paper, because: 
\begin{itemize}
\item Comprehensive analysis: an eight-page paper offers more room to deeply analyze the complexity of a flame stability problem, cover multiple angles, and explore various aspects that a four-page one might not accommodate.
\item Data and results: Sufficient space to present our results in detail and interpret results thoroughly, including
derivations, graphs, tables, and explanations of significance, which are crucial for understanding the research.
\end{itemize}

\vspace{10pt}

{\bf 4)  Colloquium Selection}
\vspace{10pt} 

\begin{itemize}

  \item{For Program Committee Members Only }

\end{itemize}

\end{@twocolumnfalse}] 


\clearpage


\section{Introduction} \addvspace{10pt}

Conjugate heat transfer (CHT) has considerable impact on the temperature and species distributions of combustion field~\cite{agostinelli2021impact, fureby2021large}, flame stabilization mechanisms~\cite{kedia2014anchoring, kedia2015blow, vance2022quantifying}, and thermoacoustic instability~\cite{kraus2018coupling}. Kraus et al.~\cite{kraus2018coupling} showed that the observed unstable thermoacoustic mode in the experiment can only be accurately predicted with carefully modeled CHT effect. CHT also allows multiplicity of flame states, according to the recent studies~\cite{kurdyumov2022flame}. Multiple different flame states with zero net heat flux between solid and fluid may be realized with the same set of parameters. Different flame states can exhibit distinctive behaviors of flame instability~\cite{dejoan2024flame}. 
These studies confirm 
that the CHT significantly complicates the flame analysis, 
and effective numerical tools are required to not only capture multiple steady-state solutions,
but also accurately predict the modal behaviors of combustion instability. 
It is noted that previous studies often leveraged the model assumption, i.e., 
infinitely high conductivity and uniform temperature in solid~\cite{dejoan2024flame}.  

In existing studies on combustion with CHT, the combustion and heat-transfer solvers are typically loosely coupled in time~\cite{kraus2018coupling, dejoan2024flame}. The steady-state solution is often obtained by either advancing both fluid and solid simulations in time or sequentially running the solvers until convergence is achieved. For combustion problems that are intrinsically stable, this segregated numerical approach can be effective. However, when combustion is unstable, the method faces significant challenges due to the vast disparity in time scales between the combustion and heat transfer processes. A smaller time-step size is generally required, and within each time step, multiple sub-iterations are necessary to enforce interface conditions, leading to substantial computational costs.

While there is increasing interest in nonlinear reactive flow simulations, linear analysis~\cite{wang2022linear, wang2022global,douglas2023flash} remains critical for efficient characterization of thermoacoustic instability, sensitivity analysis, and optimization of combustion systems. Unfortunately, the segregated algorithm becomes even more problematic when applied to eigenvalue problems, which may explain why such applications are rarely reported.

Given the needs to thoroughly study combustion instability problem with the consideration of CHT, we are motivated to develop a monolithic numerical framework that can solve the nonlinear steady-state combustion problem and 
linearized reactive Navier--Stokes equations using the same mesh and numerical schemes.  
In this framework, the fluid, solid, and interface conditions will be treated as one single mathematical framework, and the entire system is solved by a unified implicit algorithm. 
This effort will enable the understanding of 
combustion instability with CHT in frequency domain and identify the important eigenmodes, all in the same solver essentially. 

The rest of the paper is structured as follows: 
Section 2 states the problem of interest; 
Section 3 provides the detailed mathematical formulation and numerical methods; 
Boundary conditions 
and numerical setups are given in Sec. 4; 
Section 5 is devoted to the presentation of numerical results and thereby demonstrating the solver capabilities. 
The paper finishes with conclusions drawn in Sec. 6.

\section{Configuration} \addvspace{10pt}

\begin{figure}[htbp!]
 \centerline{\includegraphics[width=\textwidth]{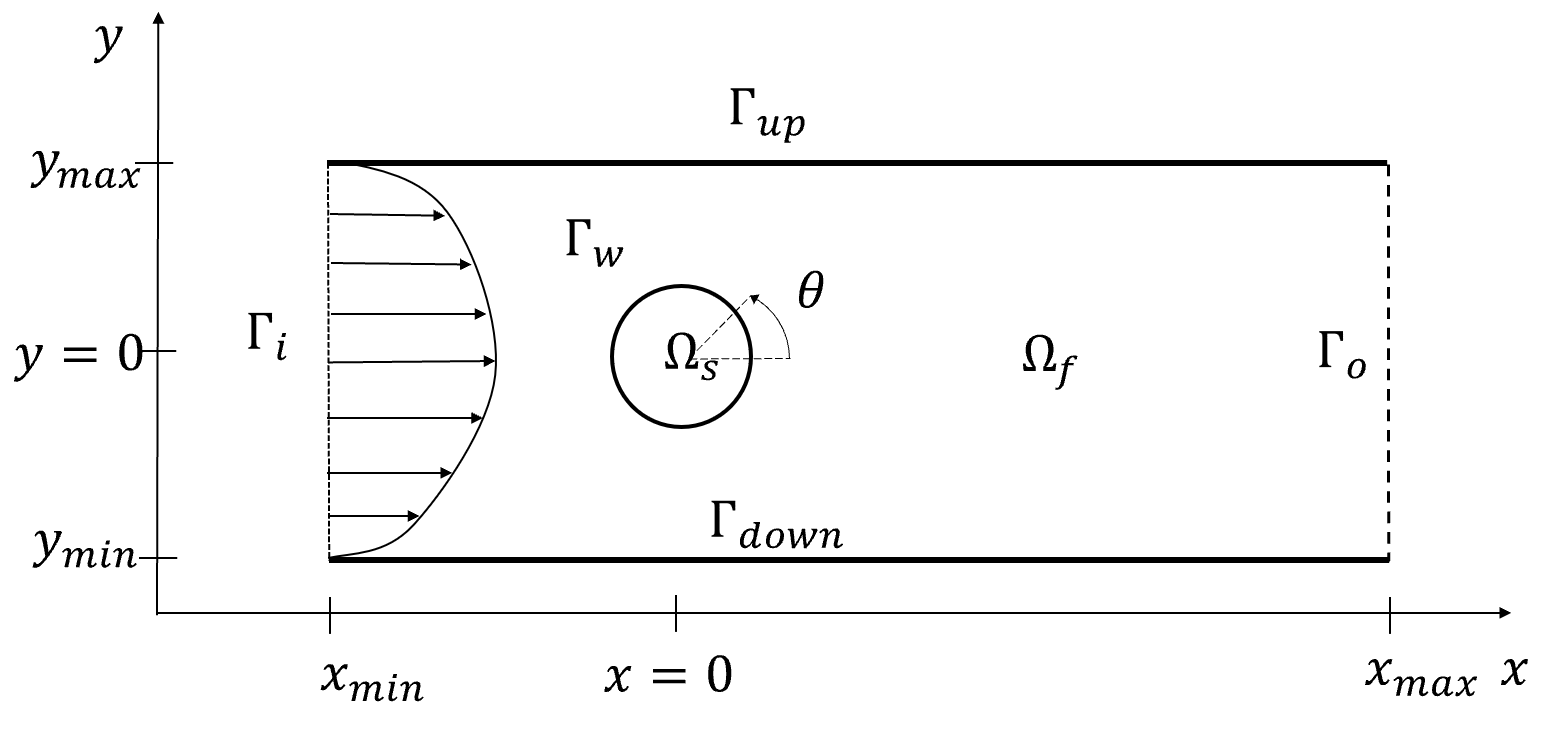}}
\caption{Illustration of the computational domain.}
\label{fig:configuration}
\end{figure}

The model combustor considered in this research is shown in Fig.~\ref{fig:configuration}. A cylinder-shape bluff body denoted as $\Omega_{s}$ is located at $x=0$ and $y=0$. The diameter of the flame holder is denoted as $D$. For the fluid part $\Omega_{f}$, the inlet $\Gamma_{i}$ and outlet $\Gamma_{o}$ boundaries are located at $x_{min} = -3.125D$ and $x_{max} = 9.375D$. The half height of the duct is $y_{max} = -y_{min} = 2.125D$. The size of the computational domain is similar with the previous research~\cite{kaiser2019impact}.

\section{Mathematical formulations} \addvspace{10pt}
\subsection{Governing equations} \addvspace{10pt}
The Navier--Stokes equations at the low-Mach number limit are employed to describe the lean premixed flame in fluid domain $\Omega_{f}$:
\begin{equation}
	\frac{\partial \rho_{f}}{\partial t}+\boldsymbol{\nabla} \boldsymbol{\cdot} (\rho_{f}\boldsymbol{u})=0 \;,
\end{equation}
\begin{equation}
	\frac{\partial (\rho_{f} \boldsymbol{u})}{\partial t}+\boldsymbol{\nabla}\boldsymbol{\cdot} (\rho_{f}\boldsymbol{u}\boldsymbol{u})=-\boldsymbol{\nabla} p + \boldsymbol{\nabla} \boldsymbol{\cdot} \boldsymbol{\tau} \;,
\end{equation}
\begin{equation}
	\frac{\partial (\rho_{f} Y_{F})}{\partial t}+\boldsymbol{\nabla} \boldsymbol{\cdot} (\rho_{f}\boldsymbol{u}Y_{F})= \boldsymbol{\nabla} \boldsymbol{\cdot} (\rho_{f} D \boldsymbol{\nabla} Y_{F}) + \dot{\omega}_{F} \;,
\end{equation}
\begin{equation}
	\frac{\partial (\rho_{f} h)}{\partial t}+\boldsymbol{\nabla}\boldsymbol{\cdot} (\rho_{f}\boldsymbol{u}h)= \boldsymbol{\nabla} \boldsymbol{\cdot} (k_{f} \boldsymbol{\nabla} T_{f}) + \dot{\omega}_{T} \;.
\end{equation}
Here, $\boldsymbol{u}=(u_{x},u_{y})$ represents the streamwise and normal velocity, $\rho_{f}$ denotes the fluid density, $p$ refers to the hydrodynamic pressure, $Y_{F}$ is the mass fraction of fuel, $T_{f}$ is the fluid temperature and $h$ is the sensible enthalpy. The thermodynamic pressure $p_{0}$ is set to be constant as $p_{0}=1$ atm. The ideal gas law is applied $p_{0}=\rho_{f} R_{g} T_{f}$ and specific gas constant $R_{g}$ is assumed to be constant for all species as $\mathrm{286.9~J/(kg\cdot K)}$. The sensible enthalpy is formulated as $h = C_{p,f}T_{f}$ and the specific heat capacity $C_{p,f}$ is also assumed to be constant for all species as $C_{p,f} = 1.004 ~\mathrm{kJ/(kg\cdot K)}$. For the diffusion terms, the viscous stress tensor is given 
as: $\boldsymbol{\tau} = \mu(\boldsymbol{\nabla}\boldsymbol{u}+\boldsymbol{\nabla}\boldsymbol{u}^{T}-2/3(\boldsymbol{\nabla}\boldsymbol{\cdot}\boldsymbol{u})\boldsymbol{I})$
and the Sutherland's law is employed to describe the viscosity $\mu = A_{s} T_{f}^{1/2}/(1+T_{c}/T_{f})$ with $A_{s} = 1.7 \times 10^{-6} ~\mathrm{kg/(m \cdot s \cdot K^{1/2})}$ and $T_{c}=170.7~\mathrm{K}$. The Prandtl number $Pr = \mu C_{p,f}/ k_{f} $ and Lewis number $ Le = k_{f} / (C_{p,f} \rho_{f} D)$ are introduced to determine the thermal conductivity $k_{f}$ and species mass diffusivity $\rho_{f} D$. In our research, the Prandtl number and Lewis number are constants as $Pr = 0.7$ and $ Le = 1.0$. For the modeling of combustion process, we choose the one-step global reaction scheme 1S$_{-}$CH4$_{-}$MP1~\cite{1SCH4MP1}. The reaction rate $\dot\omega$ is modeled by the Arrhenius law: $\dot\omega = A_{r}[X_{F}]^{n_{F}}[X_{O}]^{n_{O}}\mathrm{exp}(T_{a}/T_{f})$. Here the concentrations are related to the mass fractions by $[X_{F}]=\rho Y_{F}/W_{F}$ and $[X_{O}]=\rho Y_{O}/W_{O}$. The parameters for this chemical mechanism are listed here: Arrhenius pre-exponential factor $A_{r} = 1.1\times \mathrm{10^{7}~m^{3/2}/(s\cdot mol^{1/2})} $, activation temperature $T_{a} = 1.0065\times 10^{4}~\mathrm{K}$, exponent numbers $n_{F}=1$ and $n_{O}=0.5$ and reaction enthalpy $\Delta h_{f}^{o}=-804 ~\mathrm{kJ/mol}$. The molecular masses for fuel and oxidizer are $W_{F}=16~\mathrm{g/mol}$ and $W_{O}=32~\mathrm{g/mol}$. The combustion process is assumed to be very lean and the mass fraction of oxidizer in excess is expressed as $Y_{O}= s(Y_{F}+1/\phi-1)$, where $s$ is the stoichiometric ratio $s = 4.0$ and $\phi$ is the equivalence ratio at the inlet. 
The source terms are given by: 
$\dot\omega_{F}=-W_{F}\dot\omega$ and $\dot\omega_{T}=-\Delta h_{f}^{o} \dot\omega$. The governing equations are non-dimensionalized with inflow velocity $u_{0}$, diameter of cylinder $D$ and other flow properties at the inlet boundary, such as inlet density $\rho_{0}$, viscosity $\mu_{0}$, mass fraction $Y_{F,0}$ and temperature $T_{f,0}$. The Reynolds number is defined as $Re = (\rho_{0}u_{0}D)/\mu_{0}$ and Damk\"ohler number as $Da=(A_{r}D/u_{0})\sqrt{(\rho_{0}Y_{F,0})/W_{O}}$. In our research, we will fix the Re number to be 450 and increase the Da number during parameter continuation. Other non-dimensional parameters are the adiabatic temperature change $\Delta T=(Y_{F,0}\Delta h_{f}^{o})/(W_{F}C_{p}T_{0}) = 5.4 $ and the Zeldovich number $Ze=T_{a}\Delta T/(T_{0}(1+\Delta T)^{2})=4.4$.

For the solid part $\Omega_{s}$, the heat conduction equation is applied to describe the heat transfer process:
\begin{equation}
	\rho_{s} C_{p,s} \frac{\partial T_{s}}{\partial t} = \boldsymbol{\nabla} \boldsymbol{\cdot} (k_{s} \boldsymbol{\nabla} T_{s}) \;.
\end{equation}
Here, $T_{s}$ refers to the solid temperature, while $k_{s}$, $\rho_{s}$, and $C_{p,s}$ are 
the thermal conductivity, density and specific heat capacity of the bluff body. In this study, we choose a typical ceramic material as in the previous research~\cite{kedia2014anchoring}, with $k_{s}=1.5~\mathrm{W/(m\cdot K)}$, $\rho_{s} = 673 ~\mathrm{kg/m^{3}}$ and $C_{p,s}= 840~\mathrm{J/(kg \cdot K)}$.

At the interface $\Gamma_{w}$, we require the temperature and heat flux to be continuous in different sides: 
\begin{equation}
	T_{f} = T_{s}\;,
\end{equation}
\begin{equation}
k_{f} \boldsymbol{\nabla} T_{f} \cdot \boldsymbol{n} = k_{s} \boldsymbol{\nabla} T_{s} \cdot \boldsymbol{n} \;.
\end{equation}
Here, $\boldsymbol{n}$ refers to the normal vectors of the interface.

\subsection{Formulation of coupled problem} 
We employ the finite element method to solve the problem and apply a monolithic weak formulation to capture the coupled dynamics in the conjugate heat transfer process. The variables in fluid part and solid part are denoted as $\boldsymbol{q}_{f}$ and $\boldsymbol{q}_{s}$ with corresponding test-function denoted as $\boldsymbol{\psi}_{f}$ and $\boldsymbol{\psi}_{s}$. 
To enforce the continuity condition at the fluid-solid interface, we introduce a boundary Lagrange multiplier $\boldsymbol{\lambda}$ and its test function $\boldsymbol{\psi}_{\lambda}$ on $\Gamma_{w}$. Here, $\boldsymbol{\lambda}$ represents the heat flux on the interface between fluid and solid parts:
\begin{equation}
\label{def_lambda}
\boldsymbol{\lambda} = k_{f} \boldsymbol{\nabla} T_{f} \cdot \boldsymbol{n} = k_{s} \boldsymbol{\nabla} T_{s} \cdot \boldsymbol{n} \;.
\end{equation}
Then, we can conduct integration by parts of solid equations multiplied by its test function, which leads to:
\begin{equation}
 \begin{aligned}
   & \int_{\Omega_{s}} (\rho_{s} C_{p,s}\frac{\partial T_{s}}{\partial t} \cdot \boldsymbol{\psi}_{s})~d\Omega = \\ 
   & \int_{\Omega_{s}} (-k_{s} \boldsymbol{\nabla} T_{s} \cdot \boldsymbol{\nabla} \boldsymbol{\psi}_{s}) ~d\Omega + \int_{\Gamma_{w}} (\boldsymbol{\lambda} \cdot \boldsymbol{\psi}_{s})~d\Gamma \;.
  \end{aligned}
\end{equation}
After substituting $T_{s}$ with $\boldsymbol{q}_{s}$, we can be express it into a more compact formulation:
\begin{equation}
\label{eqn_s}
M_{s}\frac{\partial \boldsymbol{q}_{s}}{\partial t} = N_{s}(\boldsymbol{q}_{s}) + N_{s\lambda}(\boldsymbol{\lambda}) \;.
\end{equation}
Similarly, we can get the weak formulation for the fluid part:
\begin{equation}
\label{eqn_f}
M_{f}\frac{\partial \boldsymbol{q}_{f}}{\partial t} = N_{f}(\boldsymbol{q}_{f}) - N_{f\lambda}(\boldsymbol{\lambda}) \;.
\end{equation}
Here $N_{f\lambda}(\boldsymbol{\lambda}) = \int_{\Gamma_{w}}~(\boldsymbol{\lambda} \cdot \boldsymbol{\psi}_{f})~d\Gamma $ and the opposite sign before $N_{f\lambda}(\boldsymbol{\lambda})$ is due to the reversed direction of normal vectors in the solid and fluid part. 
Then, we can apply impose the temperature continuity condition on $\Gamma_{w}$ with $\boldsymbol{\psi}_{\lambda}$ as:
\begin{equation}
\int_{\Gamma_{w}} (\boldsymbol{\psi}_{\lambda} \cdot (T_{s}-T_{f})) ~d\Gamma = 0 \;.
\end{equation}
This can also be written into a compact formulation:
\begin{equation}
\label{eqn_lambda}
N_{\lambda s}(\boldsymbol{q}_{s}) - N_{\lambda f}(\boldsymbol{q}_{f}) = 0 \;.
\end{equation}
Combining Eqs.~(\ref{eqn_s}), (\ref{eqn_f}), and (\ref{eqn_lambda}), 
we can get a monolithic weak formulation for $ \boldsymbol{q}=(\boldsymbol{q}_{f},\boldsymbol{q}_{s},\boldsymbol{\lambda})^{T}$ and write it into the block-equations:
\begin{equation}
\label{eqn_coupled}
\hspace{-2mm}
\resizebox{0.42\textwidth}{!}{$
\hspace{-1mm}  \underbrace{
 \begin{pmatrix} M_{f}  & \mathbf{0} & \mathbf{0} \\  
 \mathbf{0} &  M_{s}    &   \mathbf{0}   \\
 \mathbf{0}   &   \mathbf{0}   & \mathbf{0} 
\end{pmatrix} }_{M_{\boldsymbol{q}}} 
\frac{\partial}{\partial t}
\underbrace{\begin{pmatrix}
 \boldsymbol{q}_{f}  \\  \boldsymbol{q}_{s} \\ \boldsymbol{\lambda}
\end{pmatrix}}_{\boldsymbol{q}} \hspace{-1mm}=  \hspace{-1mm}
\underbrace{
\begin{pmatrix}
N_{f}(\boldsymbol{q}_{f}) - N_{f\lambda}(\boldsymbol{\lambda}) \\ N_{s}(\boldsymbol{q}_{s}) + N_{s\lambda}(\boldsymbol{\lambda})  \\  N_{\lambda s}(\boldsymbol{q}_{s}) - N_{\lambda f}(\boldsymbol{q}_{f})
\end{pmatrix}}_{N(\boldsymbol{q})} $} \;.    \hspace{-1mm}
\end{equation}
It should be noted that the continuity condition for heat flux has been enforced implicitly with the boundary Lagrange multiplier $\boldsymbol{\lambda}$ in Eq.~(\ref{def_lambda}).

\subsection{Linear analysis method} 
The block-equations in Eq.~(\ref{eqn_coupled}) can be written as:
\begin{equation}
M_{\boldsymbol{q}} \frac{\partial \boldsymbol{q} }{\partial t} = N(\boldsymbol{q}) \;.
\end{equation}

We can decompose $\boldsymbol{q}$ into steady parts $\boldsymbol{q}_{0}$ and fluctuating parts $\boldsymbol{q'}$ as: $\boldsymbol{q} = \boldsymbol{q}_{0} + \boldsymbol{q'}$. The steady parts are referred as baseflow solutions and should satisfy:
\begin{equation}
 N(\boldsymbol{q}_{0}) = 0 \;.
\end{equation}
The linearized equations for the fluctuating parts are formulated as: 
\begin{equation}
\label{eqn_coupled_linear}
M_{\boldsymbol{q}} \frac{\partial \boldsymbol{q'} }{\partial t} = J(\boldsymbol{q}_{0}) \boldsymbol{q'}  \;.
\end{equation}
The Jacobian operator $J(\boldsymbol{q}_{0})$ for the coupled system can be expressed as block-structure:
\begin{equation}
\label{eqn_coupled}
\hspace{-2mm}
\resizebox{0.24\textwidth}{!}{$
\hspace{-1mm}  J(\boldsymbol{q}_{0})
=  \hspace{-1mm}
 \begin{pmatrix} J_{f}  & \mathbf{0} & J_{f \lambda} \\  
 \mathbf{0} &  J_{s}    &   J_{s \lambda}   \\
 J_{f \lambda}^{T}   &   J_{s \lambda}^{T}   & \mathbf{0} 
\end{pmatrix} $} \;.    \hspace{-1mm}
\end{equation}
Here, $J_{f}$ and $J_{s}$ refers to the Jacobian operator in fluid and solid equations: 
\begin{equation}
J_{f} = \frac {\partial N_{f}}{\partial \boldsymbol{q}_{f} } \bigg|_{\boldsymbol{q}_{f,0}} \quad,\quad 
J_{s} = \frac {\partial N_{s}}{\partial \boldsymbol{q}_{s} } \bigg|_{\boldsymbol{q}_{s,0}} \;.
\end{equation}
The off-diagonal blocks $J_{f \lambda}$ and $J_{s \lambda}$ are written as:
\begin{equation}
J_{f \lambda} = \int_{\Gamma_{w}} ( - \boldsymbol{\lambda'} \cdot \boldsymbol{\psi}_{f})~d\Gamma \;,
\end{equation}
\begin{equation}
J_{s \lambda} = \int_{\Gamma_{w}} ( \boldsymbol{\lambda'} \cdot \boldsymbol{\psi}_{s})~d\Gamma \;.
\end{equation}

The above linearized formulations in Eq.~(\ref{eqn_coupled_linear}) can be employed to conduct linear stability analysis and resolvent analysis respectively. 

For the linear stability analysis, the state vector $\boldsymbol{q'}$ is expanded into normal mode $\boldsymbol{\hat{q}}$: $\boldsymbol{q'}(x,y,t) = \boldsymbol{\hat{q}}(x,y) \mathrm{exp}((\omega_{r}+ i \omega_{i})t)$ and corresponding eigenvalue problem is expressed as:
\begin{equation}
J(\boldsymbol{q}_{0}) \boldsymbol{\hat{q}} =  (\omega_{r}+ i \omega_{i}) M_{\boldsymbol{q}} \boldsymbol{\hat{q}} \;.
\end{equation}
Here, $\omega_{r}$ and $\omega_{i}$ refer to the growth rate and frequency of this eigenmode. In case $\omega_{r}$ is greater than zero, this eigenmode is referred to be unstable, otherwise stable. If all the eigenmodes are stable, we can further conduct resolvent analysis to investigate the amplification behavior of this linear system. 

In the resolvent analysis, we apply Fourier transform to the fluctuating parts and consider an external force $\boldsymbol{\hat{f}}$ in the right hand side of Eq.~(\ref{eqn_coupled_linear}):
\begin{equation}
  i \omega_{i} M_{\boldsymbol{q}} \boldsymbol{\hat{q}} = J(\boldsymbol{q}_{0}) \boldsymbol{\hat{q}} +  P \boldsymbol{\hat{f}} \;.
\end{equation}
Then, we can define the resolvent operator $R_{\boldsymbol{q}}$ with the input-output relation:
\begin{equation}
 \boldsymbol{\hat{q}} =
 \underbrace{
 (i \omega_{i} M_{\boldsymbol{q}} - J(\boldsymbol{q}_{0})) ^{-1} 
 }_{R_{\boldsymbol{q}}} P \boldsymbol{\hat{f}} \;.
\end{equation}
Here, the operator $P$ is applied to relate the dimension of forcing to responding fields. To measure the amplification effects of responding field $\boldsymbol{\hat{q}}$ to forcing field $\boldsymbol{\hat{f}}$, we introduce to energy normals $W_{\boldsymbol{q}}$ and $W_{\boldsymbol{f}}$:
\begin{equation}
||\boldsymbol{\hat{q}}||^{2} = \boldsymbol{\hat{q}}^{H}W_{\boldsymbol{q}}\boldsymbol{\hat{q}} \quad,\quad
||\boldsymbol{\hat{f}}||^{2} = \boldsymbol{\hat{f}}^{H}W_{\boldsymbol{f}}\boldsymbol{\hat{f}} \;.
\end{equation}
In our research, we consider external forcing field $\boldsymbol{\hat{f}}=(\hat{f}_{x},\hat{f}_{y})^{T}$ in the momentum equations of fluid domain and focus on the reaction rate in the responding field to define the energy norms as:
\begin{equation}
||\boldsymbol{\hat{q}}||^{2} = \int_{\Omega_{f}} (\hat{\dot\omega})^{2} ~d\Omega \;,
\end{equation}
\begin{equation}
||\boldsymbol{\hat{f}}||^{2} = \int_{\Omega_{f}} (\hat{f}_{x}^{2}+\hat{f}_{y}^{2})~d\Omega \;.
\end{equation}
The maximum value of amplification is expressed as:
\begin{equation}
\sigma^{2} = \max_{\boldsymbol{\hat{f}}} \frac{||\boldsymbol{\hat{q}}||^{2}}{||\boldsymbol{\hat{f}}||^{2}} = 
\max_{\boldsymbol{\hat{f}}} \frac{\boldsymbol{\hat{f}}^{H}P^{H}R_{\boldsymbol{q}}^{H}W_{\boldsymbol{q}}R_{\boldsymbol{q}}P\boldsymbol{\hat{f}}}{\boldsymbol{\hat{f}}^{H}W_{\boldsymbol{f}}\boldsymbol{\hat{f}}} \;.
\end{equation}
Here, the value of $\sigma^{2}$ measures the optimal amplification of this linear system to any external forcing.

\section{Numerical details} \addvspace{10pt}

\begin{table}[ht] \footnotesize
\newcommand{\tabincell}[2]{\begin{tabular}{@{}#1@{}}#2\end{tabular}}
\centering
\begin{tabular}{lc}
  \hline
    Boundary & Constraints   \\
  \hline
      $\Gamma_{i}$  & \tabincell{l}{\rule{0pt}{8pt}$u_{x}=1-(y/y_{max})^{6.75}$,~~ $u_{y}=0$,\\ \rule{0pt}{8pt}$Y_{F}=1$,~~$T_{f}=1$,~~$\phi = 0.75$}\\
  \hline
      $\Gamma_{up}$,~$\Gamma_{down}$ & \tabincell{l}{\rule{0pt}{8pt}$u_{x}=u_{y}=0$, \\ \rule{0pt}{8pt}$\nabla Y_{F} \cdot \boldsymbol{n}=0$,~~$T_{f}=1$}\\
  \hline
     $\Gamma_{w}$  & \tabincell{l}{\rule{0pt}{8pt}$u_{x}=u_{y}=0$,~~$\nabla Y_{F} \cdot \boldsymbol{n}=0$
     \\ \rule{0pt}{8pt}$T_{f}=T_{s}$,~~$k_{f} \boldsymbol{\nabla} T_{f} \cdot \boldsymbol{n} = k_{s} \boldsymbol{\nabla} T_{s} \cdot \boldsymbol{n}$}\\
  \hline
     $\Gamma_{o}$ & \tabincell{l}{ \rule{0pt}{8pt}$(-p\boldsymbol{I}+\mu \nabla\boldsymbol{u})\cdot\boldsymbol{n}=0$,\\ \rule{0pt}{8pt}$\nabla Y_{F} \cdot \boldsymbol{n} = \nabla T_{f} \cdot \boldsymbol{n}=0$}\\
  \hline
\end{tabular}
\caption{Boundary conditions for the baseflow calculation.}
\label{tab:boundary conditions for baseflow}
\end{table}

\begin{table}[ht] \footnotesize
\newcommand{\tabincell}[2]{\begin{tabular}{@{}#1@{}}#2\end{tabular}}
\centering
\begin{tabular}{lc}
  \hline
    Boundary & Constraints   \\
  \hline
      $\Gamma_{i}$  & \rule{0pt}{8pt}$\hat{u}_{x}=\hat{u}_{y}=\hat{Y}_{F}=\hat{T}_{f}=0$\\
  \hline
      $\Gamma_{up}$,~$\Gamma_{down}$ & \rule{0pt}{8pt}$\hat{u}_{x}=\hat{u}_{y}=\hat{T}_{f}=0$,~~$
      \boldsymbol{\nabla} \hat{Y}_{F} \cdot \boldsymbol{n} =0$\\
  \hline
     $\Gamma_{w}$  & \tabincell{l}{\rule{0pt}{8pt}$\hat{u}_{x}=\hat{u}_{y}=0$,~~$
      \boldsymbol{\nabla} \hat{Y}_{F} \cdot \boldsymbol{n} =0$
     \\ \rule{0pt}{8pt}$\hat{T}_{f}=\hat{T}_{s}$,~~$k_{f} \boldsymbol{\nabla} \hat{T}_{f} \cdot \boldsymbol{n} = k_{s} \boldsymbol{\nabla} \hat{T}_{s} \cdot \boldsymbol{n}$}\\
  \hline
     $\Gamma_{o}$ & \tabincell{l}{ \rule{0pt}{8pt}$(-\hat{p}\boldsymbol{I}+\mu \nabla\boldsymbol{\hat{u}})\cdot\boldsymbol{n}=0$,\\ \rule{0pt}{8pt}$\nabla \hat{Y}_{F} \cdot \boldsymbol{n} = \nabla \hat{T}_{f} \cdot \boldsymbol{n}=0$}\\
  \hline
\end{tabular}
\caption{Boundary conditions for linear analysis.}
\label{tab:boundary conditions for linear analysis}
\end{table}

The boundary conditions for the calculation of baseflow solutions and linear analysis are listed in Tables~\ref{tab:boundary conditions for baseflow} and 
\ref{tab:boundary conditions for linear analysis}. The governing equations combined with boundary conditions were discretized with the open-source software FreeFem++~\cite{hecht2012new} and results are gathered with the open-source drivers StabFem~\cite{fabre2018practical}. All variables were discretized with P2 element, except for the hydrodynamic pressure with P1 element. The total mesh comprises about 160,000 triangles. 
The flame is resolved with at 10 mesh points. 
The baseflow solutions were calculated with Newton iteration method. 
We employed the Moore-Penrose method to conduct the numerical parameter continuation with Da number increasing to evaluate the multiple baseflow solutions, following the previous research~\cite{douglas2021nonlinear}. 
For each baseflow solutions, we conducted linear stability analysis and resolvent analysis. 
The general eigenvalue problem in linear stability analysis is solved with solved using the Krylov--Schur method. 
The optimization problem in resolvent analysis is first transformed into a singular value problem and then a general eigenvalue problem, 
as in the previous study~\cite{sipp2013characterization}. 
The non-dimensional frequency resolution is 0.1. 
We used open-source libraries PETSc~\cite{petsc-efficient} and SLEPc~\cite{Hernandez:2005:SSF} to 
perform the linear algebra tasks and eigenvalue problems. 
The codes for calculating steady conjugate heat transfer problem are validated in Appendix A.

\section{Results and Discussion} \addvspace{10pt}

\subsection{Baseflow Solutions}\addvspace{10pt}
\begin{figure*}[htbp!]
 \centerline{\includegraphics[width=\textwidth]{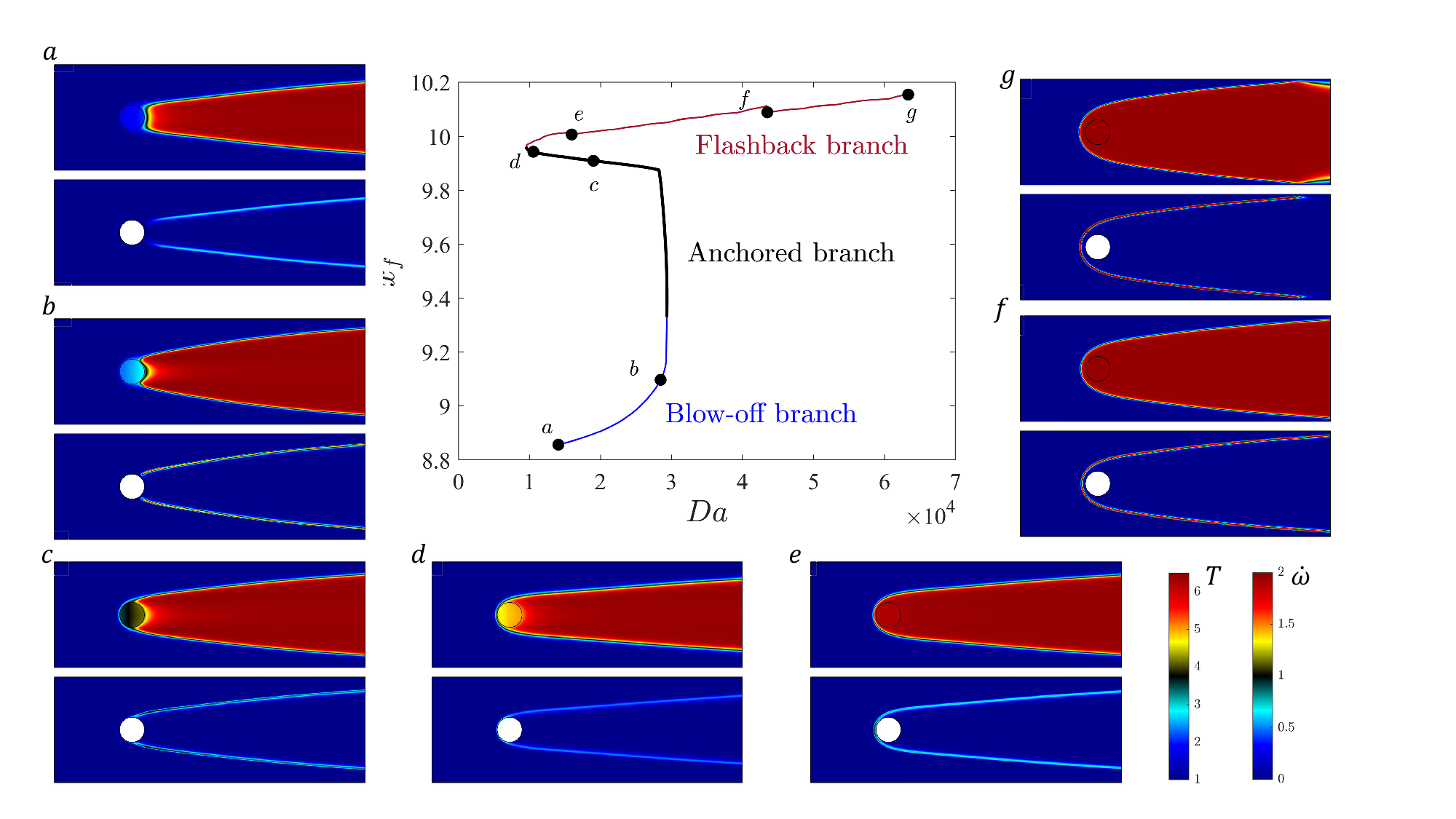}}
\caption{Bifurcation diagram showing the evolution of the length of the flame ($x_{f}$) with respect to Da number. Visualization of the temperature and reaction rate contours at different points along the bifurcation curve.}
\label{fig:BaseFlow_continuation}
\end{figure*}

\begin{figure*}[ht]
     \begin{subfigure}[b]{0.4\textwidth}
         \centering
         \includegraphics[width=\textwidth]{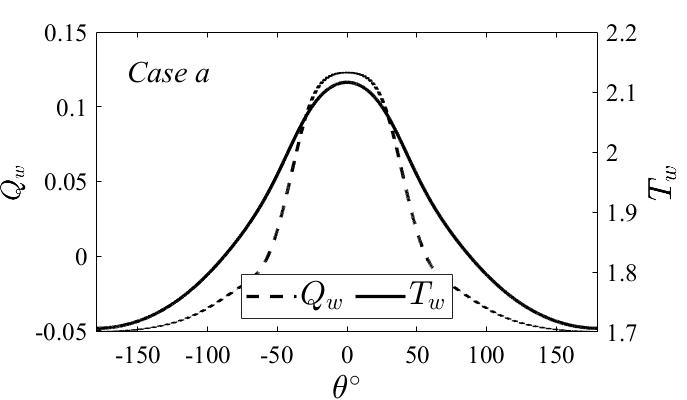}
     \end{subfigure}
     \begin{subfigure}[b]{0.4\textwidth}
         \centering
         \includegraphics[width=\textwidth]{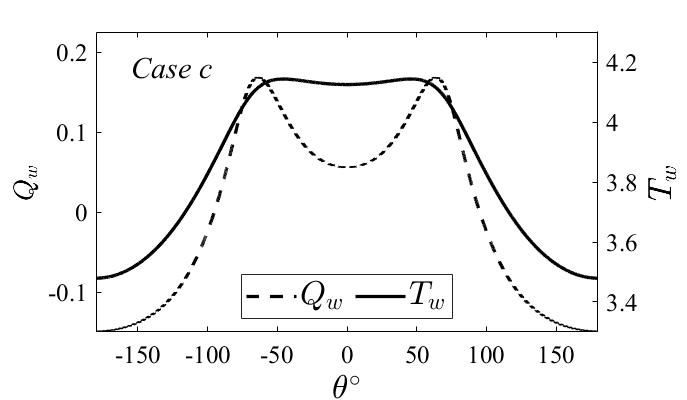}
     \end{subfigure}
     \begin{subfigure}[b]{0.4\textwidth}
         \centering
         \includegraphics[width=\textwidth]{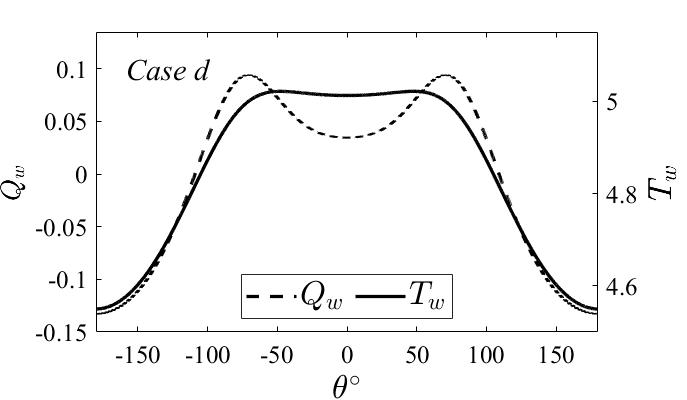}
     \end{subfigure}
     \begin{subfigure}[b]{0.4\textwidth}
         \centering
         \includegraphics[width=\textwidth]{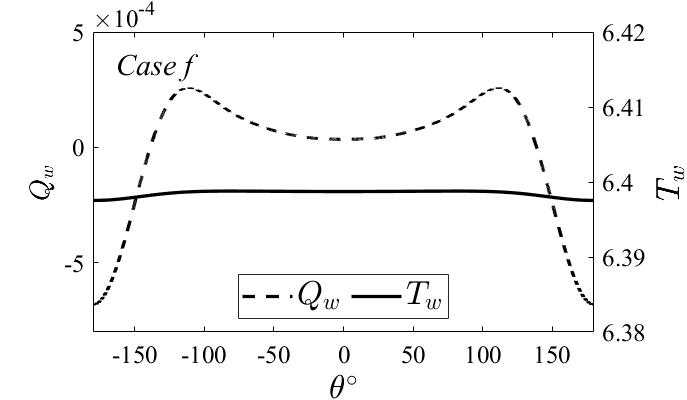}
     \end{subfigure}
\caption{Heat flux and temperature profiles on the cylinder surface for Cases $a$, $c$, $d$, and $f$ in Fig.~\ref{fig:BaseFlow_continuation}.}     
\label{fig:surface}     
\end{figure*}

The results for numerical parameter continuation of baseflow solutions along the Da number are shown in Fig.~\ref{fig:BaseFlow_continuation}. We evaluate the position of the left point of the flame: $x_{f}=\min_{x} (0.05\max(\dot{\omega}))$ while increasing the Da number. 
Here, we mostly focus on the ignited flame and corresponding Da number range. The bifurcation diagram demonstrates there exist multiple steady states during this range, similar to 
the previous study~\cite{dejoan2024flame}. 
From the temperature and reaction rate fields, we can classify the bifurcation curve and these baseflow solutions into three different branches: 
(i)   blow-off branch (Cases $a$ and $b$), 
(ii)  anchored branch (Cases $c$ and $d$), and 
(iii) flashback branch (Cases $e$, $f$, and $g$). 
The flame can be anchored to the cylinder only within a limited Da range shown in thick black line. For cases of Da number below this range, the flame cannot be anchored to the cylinder because the timescale of chemical reaction is slower than that of flow. However, for cases of higher Da number, the flashback of flame is observed. This phenomenon can be attributed to the heat conduction across the cylinder. The heat released by the flame first increases the temperature on the back side of the cylinder. Then the heat is transferred to the front side across the cylinder to heat up the upstream unburnt gas. As the chemical reaction in the front of the cylinder becomes more active, the flame front moves upstream to yield a flashback situation. This phenomenon can also be investigated from the heat flux $Q_{w}$ and temperature $T_{w}$ distribution along the cylinder surface, as shown in Fig.~\ref{fig:surface} for Cases $a$, $c$, $d$, and $f$ labeled in Fig.~\ref{fig:BaseFlow_continuation}. For Case $a$, the single-peak patterns are observed on the temperature and heat-flux curves for the lifted flame. As the flame is attached to the cylinder surface in Cases $c$ and $d$, two peaks are present in the heat-flux curve corresponding to the two flame anchoring points. As the Da number continues to grow, the flashback flame appears in Case $f$. 
The cylinder is fully surrounded by the burnt gas and the solid surface becomes approximately isothermal  
with negligible heat transfer between the flame and cylinder. 

\subsection{Linear stability analysis}\addvspace{10pt}

\begin{figure*}[t]
     \begin{subfigure}[b]{0.4\textwidth}
         \centering
         \includegraphics[width=\textwidth]{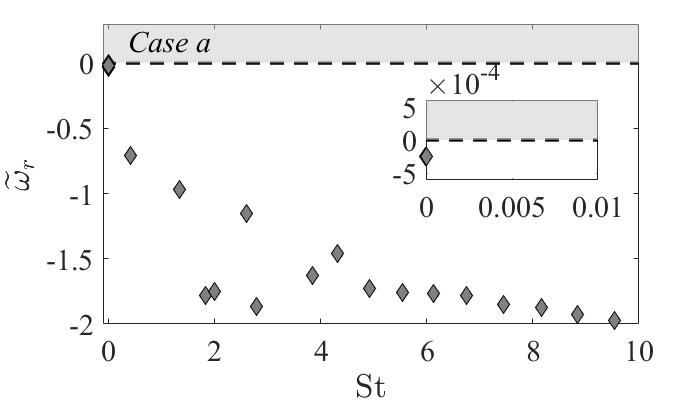}
     \end{subfigure}
     \begin{subfigure}[b]{0.4\textwidth}
         \centering
         \includegraphics[width=\textwidth]{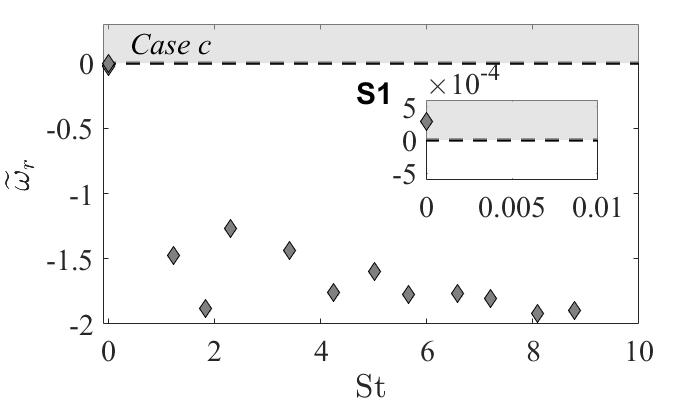}
     \end{subfigure}
     \begin{subfigure}[b]{0.4\textwidth}
         \centering
         \includegraphics[width=\textwidth]{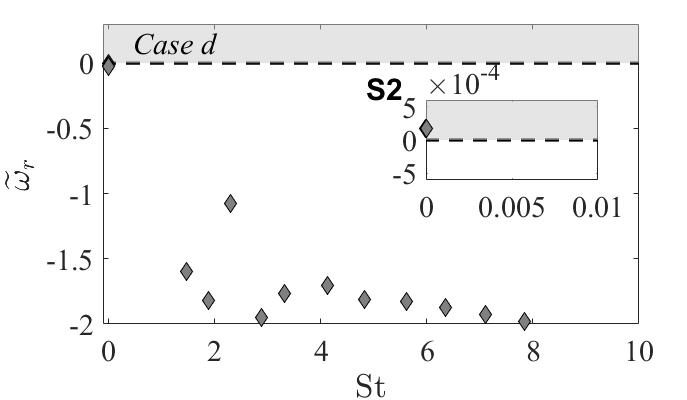}
     \end{subfigure}
     \begin{subfigure}[b]{0.4\textwidth}
         \centering
         \includegraphics[width=\textwidth]{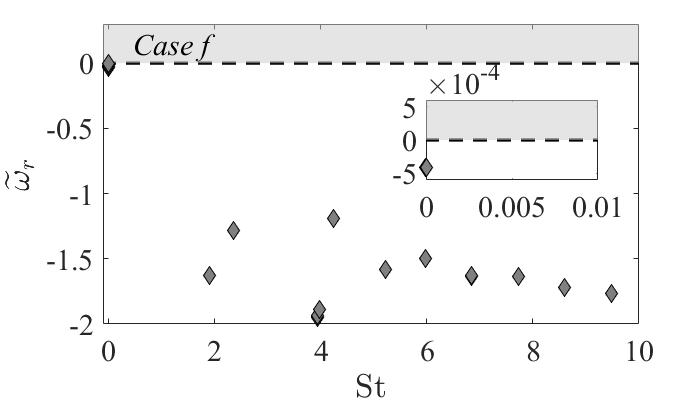}
     \end{subfigure}
\caption{Eigenspectra for Cases $a$, $c$, $d$, and $f$ in Fig.~\ref{fig:BaseFlow_continuation}. The region near the origin is magnified in the insets.}          
\label{fig:Eigenspectrum}     
\end{figure*}

\begin{figure}[htb]
     \begin{subfigure}[b]{0.865\textwidth}
         \centering
         \includegraphics[width=\textwidth]{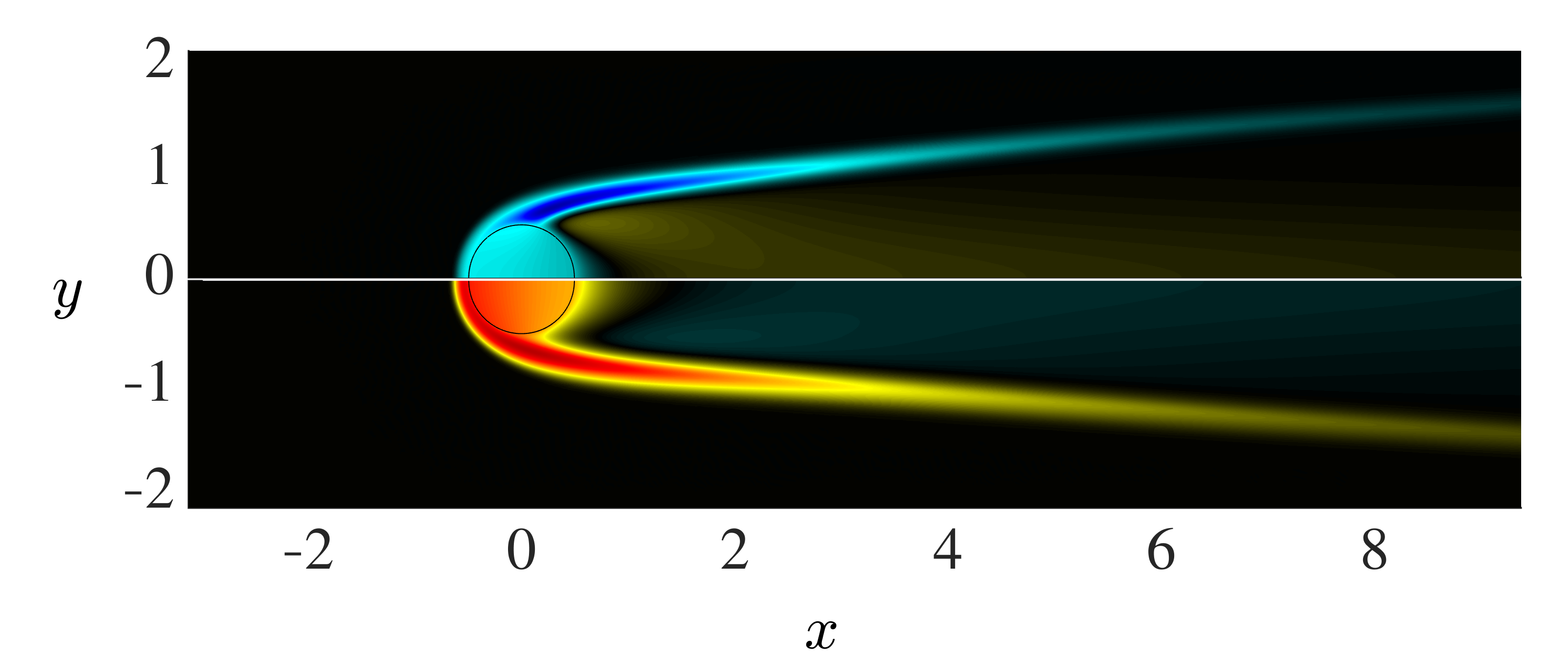}
     \end{subfigure}
     \begin{subfigure}[b]{0.11\textwidth}
         \centering
         \includegraphics[width=\textwidth]{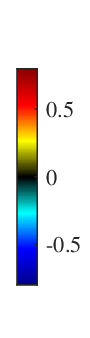}
     \end{subfigure}
\caption{Real part of the temperature field for eigenmodes S1 (top half) in Case $c$ and S2 (bottom half) in Case $d$.}          
\label{fig:eigenmodesT}     
\end{figure}

\begin{figure}[htb]
     \begin{subfigure}[b]{0.865\textwidth}
         \centering
         \includegraphics[width=\textwidth]{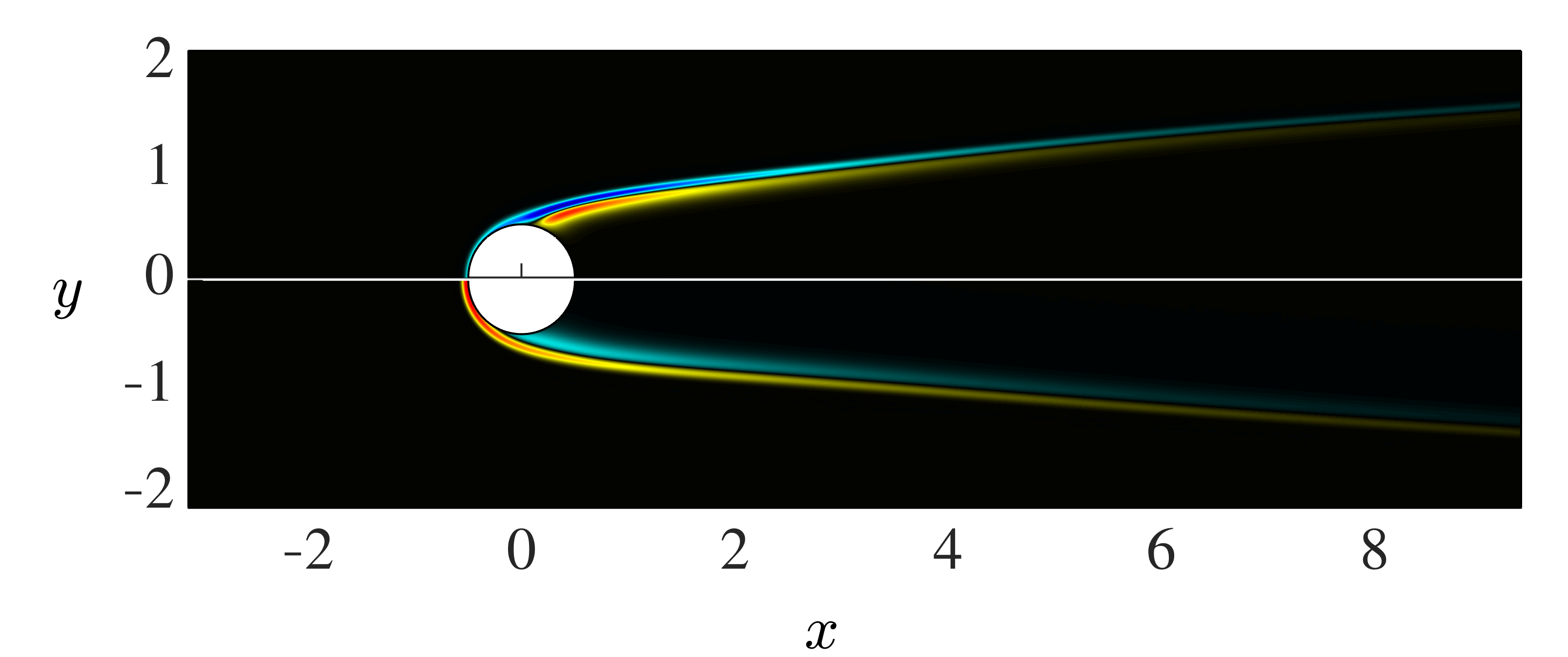}
     \end{subfigure}
     \begin{subfigure}[b]{0.1\textwidth}
         \centering
         \includegraphics[width=\textwidth]{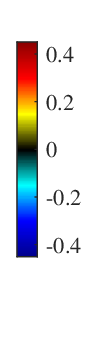}
     \end{subfigure}
\caption{Real part of the reaction rate field for eigenmodes S1 (top half) in Case $c$ and S2 (bottom half) in Case $d$.}          
\label{fig:eigenmodesQ}     
\end{figure}

The linear stability analysis is carried out to study the four cases shown in Fig.~\ref{fig:surface} and the eigenspectra are exhibited in Fig.~\ref{fig:Eigenspectrum}. The non-dimensional frequency and growth rate are denoted as $\mathrm{St} = \omega_{i} D/u_{0}$ and $\widetilde{\omega}_{r}= \omega_{r} D/u_{0}$. 
For Cases $a$ and $f$, all the eigenmodes are stable. 
However, one non-oscillating unstable eigenmode is found in each Cases $c$ and $d$ of the anchored flame branch, 
namely the S1 and S2 modes in Fig.~\ref{fig:Eigenspectrum}. 
We present the temperature and reaction rate fields of these two eigenmodes in Figs.~\ref{fig:eigenmodesT} and 
\ref{fig:eigenmodesQ}. These two eigenmodes are mostly supported along the flame sheet, 
but also feature 
significant temperature fluctuations inside the cylinder. This implies that heat conduction is dominant in frequencies much lower than the fluid flows. The differences between the S1 and S2 are related to the anchoring-point locations in the two baseflows. 

\subsection{Resolvent analysis}\addvspace{10pt}
The resolvent analysis is conducted to further investigate the coupled dynamics of flame with CHT. 
Note that, even though two unstable eigenmodes have been identified in Cases $c$ and $d$, 
the flow may still be influenced by the amplification of external forcing, 
especially when 
the characteristic frequencies of solid and fluid parts are different. 

\begin{figure*}[ht]
     \begin{subfigure}[b]{0.4\textwidth}
         \centering
         \includegraphics[width=\textwidth]{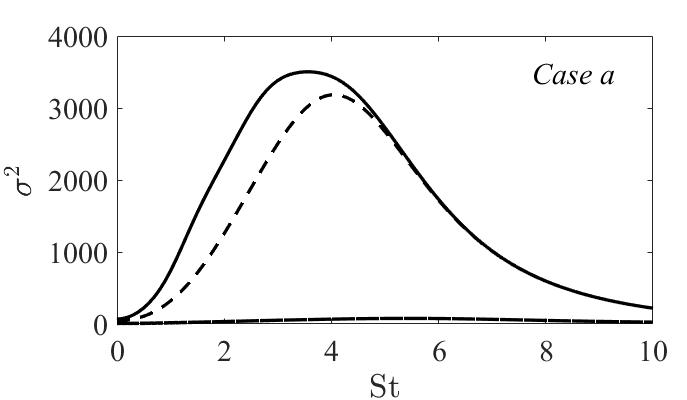}
     \end{subfigure}
     \begin{subfigure}[b]{0.4\textwidth}
         \centering
         \includegraphics[width=\textwidth]{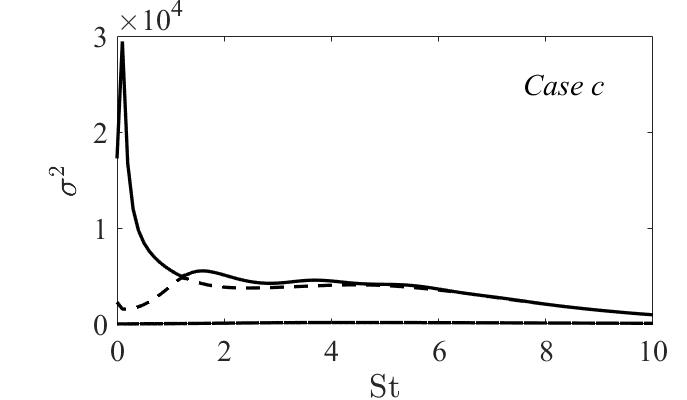}
     \end{subfigure}
     \begin{subfigure}[b]{0.4\textwidth}
         \centering
         \includegraphics[width=\textwidth]{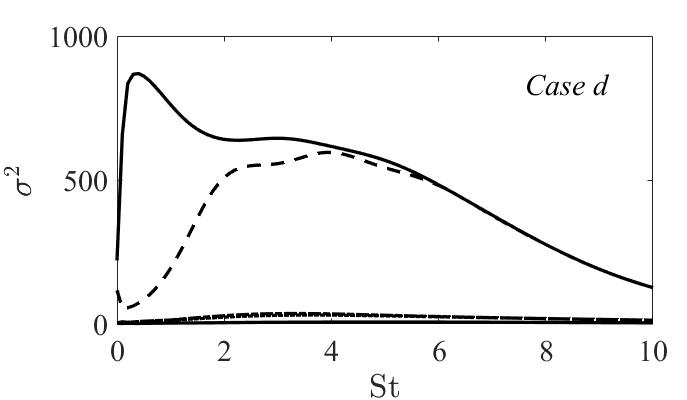}
     \end{subfigure}
     \begin{subfigure}[b]{0.4\textwidth}
         \centering
         \includegraphics[width=\textwidth]{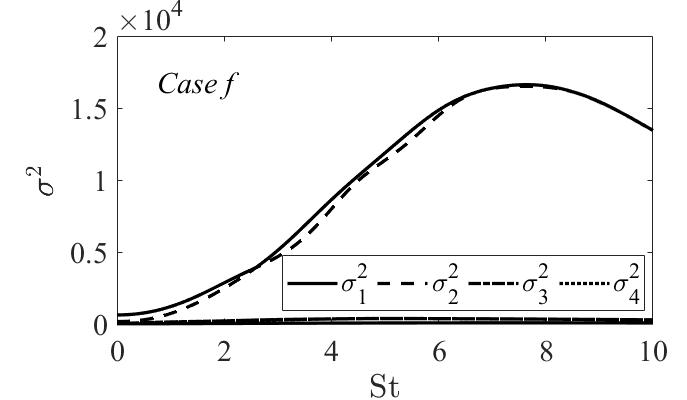}
     \end{subfigure}
    \caption{Gain curves of the first four dominant resolvent  modes for  Cases $a$, $c$, $d$, and $f$ in Figure~\ref{fig:BaseFlow_continuation}.}
\label{fig:resolvent_gain} 
\end{figure*}

\begin{figure}[htb]
     \begin{subfigure}[b]{0.865\textwidth}
         \centering
         \includegraphics[width=\textwidth]{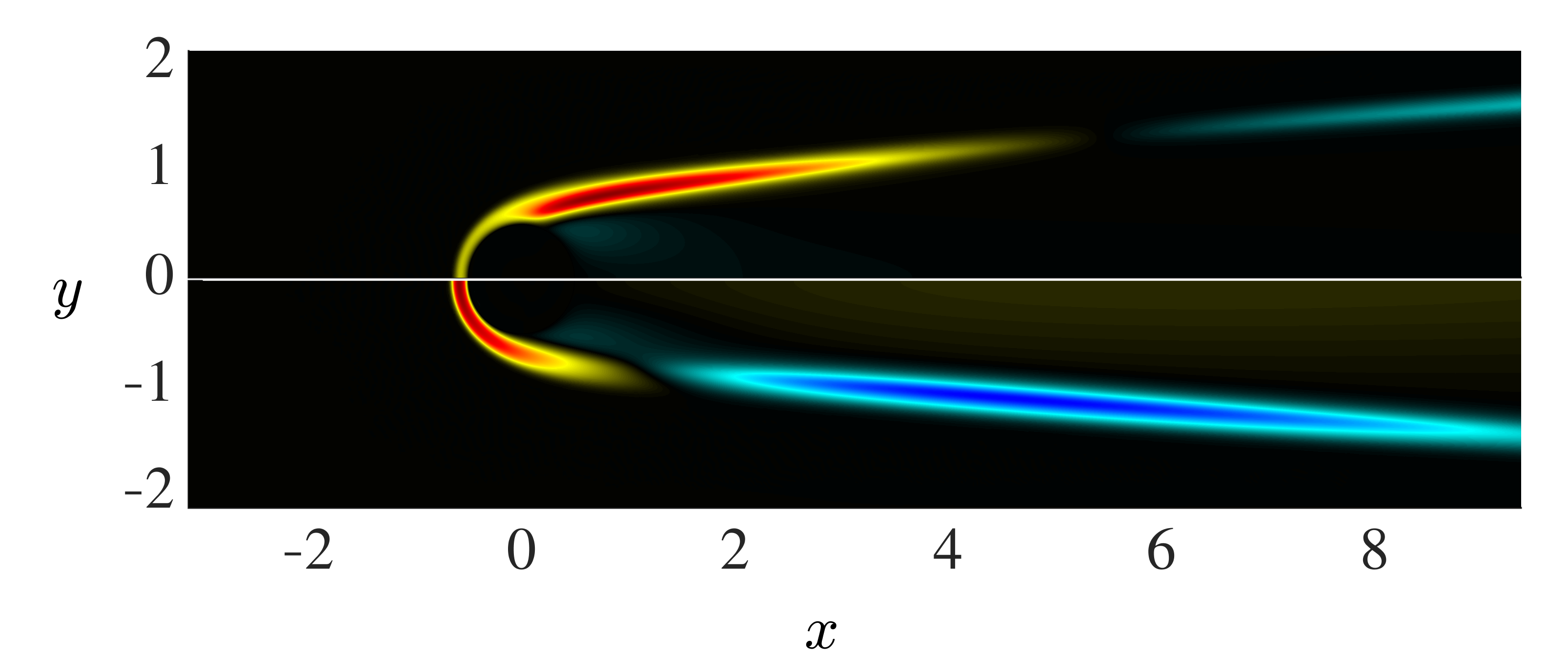}
     \end{subfigure}
     \begin{subfigure}[b]{0.1\textwidth}
         \centering
         \includegraphics[width=\textwidth]{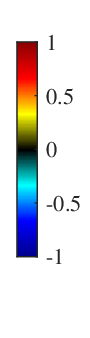}
     \end{subfigure}
\caption{Real part of the responding temperature field at $\mathrm{St}=0.1$ for Case $c$ (top half) and $0.4$ for Case $d$ (bottom half).}
\label{fig:resolv1}     
\end{figure}

\begin{figure}[htb]
     \begin{subfigure}[b]{0.865\textwidth}
         \centering
         \includegraphics[width=\textwidth]{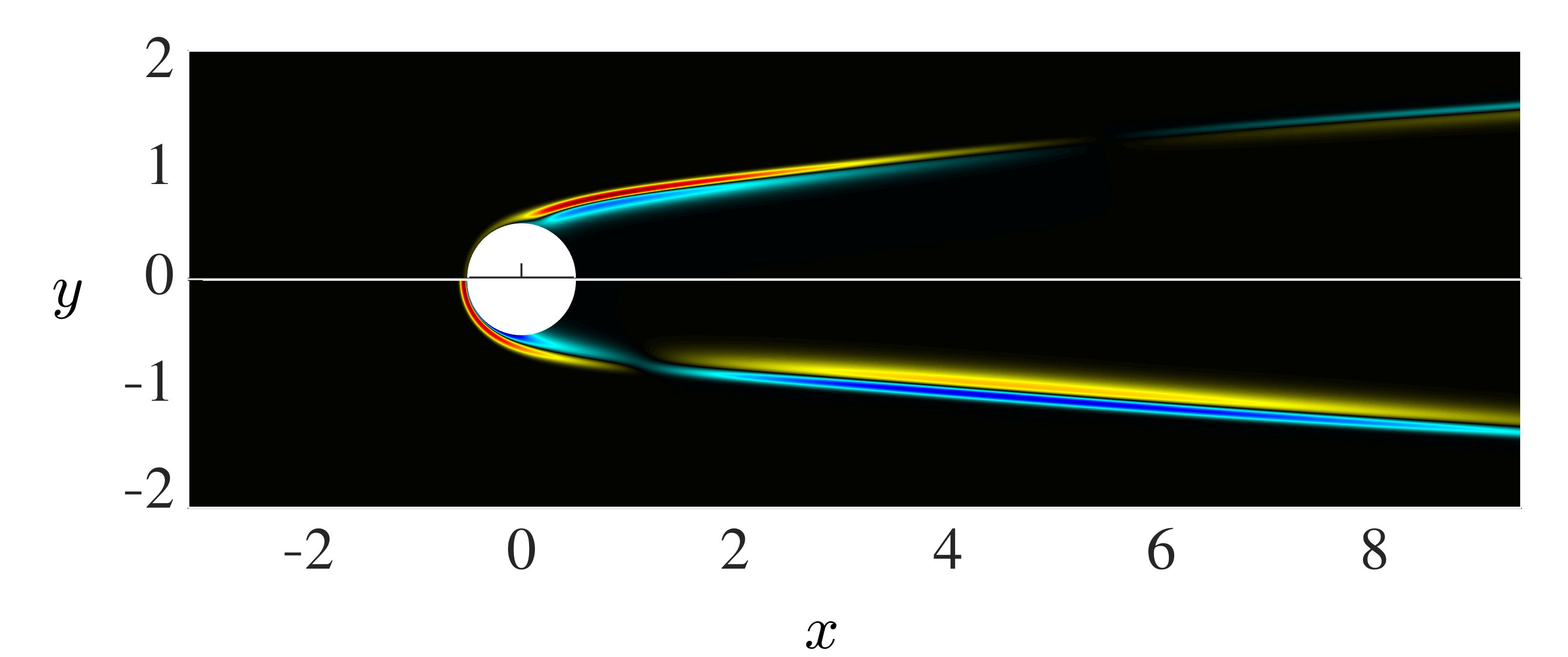}
     \end{subfigure}
     \begin{subfigure}[b]{0.1\textwidth}
         \centering
         \includegraphics[width=\textwidth]{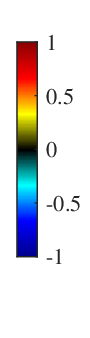}
     \end{subfigure}
\caption{Real part of the responding reaction rate field at $\mathrm{St}=0.1$ for Case $c$ (top half) and $0.4$ for Case $d$ (bottom half).}  
\label{fig:resolv2}     
\end{figure}

The four leading resolvent gain curves of Cases $a$, $c$, $d$, and $f$ are presented in Fig.~\ref{fig:resolvent_gain}. An intense peak in gain curve of $\sigma_{1}^{2}$ is identified in Case $c$ at $\mathrm{St} \approx 0.1$. This is suspected to be the resonance phenomenon due to the existence of eigenmodes close to the frequency of forcing. However, for Case $d$, there exists another peak near $\mathrm{St} \approx 0.4$. Here, we display the responding fields for temperature and reaction rate in Figs.~\ref{fig:resolv1} and \ref{fig:resolv2}. For the temperature fields, the temperature fluctuations are much lower in the solid part than the corresponding unstable eigenmodes, implying 
that the total heat fluctuations released by the flame will peak 
when the heat losses due to conjugate heat transfer are minimized. For Cases $a$ and $f$, the peaks in gain curves are suspected to be the pseudo-resonance of shear flows~\cite{symon2018non}.

\section{Conclusions}\addvspace{10pt}
The coupling between an anchored flame and a thermal conductive cylinder has been studied. Multiple steady states have been identified. The bifurcation diagram indicates the existence of blow-off, anchored flame, and flashback branches as Damk\"{o}hler number increases. 
The results of linear stability show that each anchored flame is featured with one unstable non-oscillating eigenmode. 
The optimal responding modes in resolvent analysis imply that the heat fluctuations released by anchored flame will be maximized with minimum 
heat transfer between fluid and solid. 
Further research can be conducted to attribute the amplification to solid and fluid part, allowing the 
design of stable combustion system in the future.

\begin{figure}[htbp!]
    \centering
    \begin{subfigure}{1.0\textwidth}
        \centering
        \centerline{
        \begin{overpic}[width=1\textwidth]{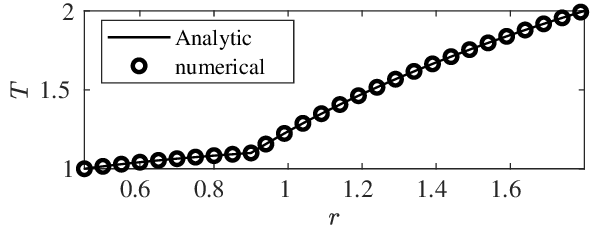}
        \end{overpic}
        }
        \phantomsubcaption        
    \end{subfigure}
\caption{The temperature profile in the validation case.}
\label{fig:validation_case}
\end{figure}

\acknowledgement{Appendix A: Validation of the codes} \addvspace{10pt}
Our codes have been validated with an axisymmetric conjugate heat transfer problem of rotating flows between two coannular cylinders with the inner cylinder composed of thermal conductive solids, as reported in the previous study~\cite{kang2009dns}. The ratio of thermal conductivity between the solid and fluid part is set to be 9 and the non-dimensional angular velocity of the outer cylinder is 1. 
The temperature field 
along the radial direction is 
compared to the analytic solutions provided in the literature, as shown in Fig.~\ref{fig:validation_case}. 
The result confirms 
the accuracy of our codes.

\acknowledgement{Declaration of competing interest} \addvspace{10pt}
The authors have no conflicts to disclose.

\acknowledgement{Acknowledgments} \addvspace{10pt}
LC acknowledges Dr. C.M. Douglas from Duke University and Dr. P. Jolivet from Sorbonne Universit\'{e} on help with numerical algorithm and parallel computation with FreeFem++, including the open-source ff-bifbox package (https://github.com/cmd8/ff-bifbox). 
YL acknowledges the fundings from NSFC (No. 12472298) and CAS (No. 025GJHZ2022112FN). 
WLC acknowledges the funding from CAS (No. 2023VMC0022).


 \footnotesize
 \baselineskip 9pt


\bibliographystyle{pci}
\bibliography{PCI_LaTeX}


\newpage

\small
\baselineskip 10pt



\end{document}